\documentstyle[12pt]{article}

\begin{document}
\newcommand{\pl}[1]{Phys.\ Lett.\ {\bf #1}\ }
\newcommand{\npb}[1]{Nucl.\ Phys.\ {\bf B#1}\ }
\newcommand{\prd}[1]{Phys.\ Rev.\ {\bf D#1}\ }
\newcommand{\prl}[1]{Phys.\ Rev.\ Lett.\ {\bf #1}\ }
\newcommand{\hepph}[1]{{\tt hep-ph/#1}}
\newcommand{\hepth}[1]{{\tt hep-th/#1}}

\newcommand{\drawsquare}[2]{\hbox{%
\rule{#2pt}{#1pt}\hskip-#2pt
\rule{#1pt}{#2pt}\hskip-#1pt
\rule[#1pt]{#1pt}{#2pt}}\rule[#1pt]{#2pt}{#2pt}\hskip-#2pt
\rule{#2pt}{#1pt}}

\newcommand{\Yfund}{\raisebox{-.5pt}{\drawsquare{6.5}{0.4}}}
\newcommand{\Ysymm}{\raisebox{-.5pt}{\drawsquare{6.5}{0.4}}\hskip-0.4pt%
        \raisebox{-.5pt}{\drawsquare{6.5}{0.4}}}
\newcommand{\Yasymm}{\raisebox{-3.5pt}{\drawsquare{6.5}{0.4}}\hskip-6.9pt%
        \raisebox{3pt}{\drawsquare{6.5}{0.4}}}

\begin{titlepage}
\begin{center}
{\hbox to\hsize{hep-th/9605108 \hfill  MIT-CTP-2532}}

\bigskip

\bigskip

{\Large \bf  More Dynamical Supersymmetry Breaking
\footnotemark[1]      } \\

\bigskip

\bigskip

{\bf Csaba Cs\'aki, Lisa Randall\footnotemark[2] and Witold Skiba}\\

\smallskip

{ \small \it Center for Theoretical Physics

Laboratory for Nuclear Science and Department of Physics

Massachusetts Institute of Technology

Cambridge, MA 02139, USA }

 \bigskip

\vspace{1cm}

{\bf Abstract}\\
\end{center}
In this paper we introduce a new class of theories which dynamically
break supersymmetry based on the gauge group $SU(n)\times SU(3) \times U(1)$
for even $n$. These theories are interesting in that no dynamical
superpotential is generated  in the absence of perturbations.
For the example $SU(4) \times SU(3) \times U(1)$ we explicitly
demonstrate that all flat directions can be lifted through
a renormalizable superpotential and that supersymmetry is dynamically broken.  
We derive the exact superpotential for this theory, which  exhibits new
and interesting dynamical phenomena. For example, modifications to classical
constraints can be field dependent. We also consider the generalization to
$SU(n) \times SU(3) \times U(1) $ models (with even $n>4$). We present
a renormalizable superpotential which lifts all flat directions. Because
$SU(3)$ is not confining in the absence of perturbations, the analysis of
supersymmetry breaking is very different in these theories from the $n=4$
example. When the $SU(n)$ gauge group confines, the Yukawa couplings drive the 
$SU(3)$ theory into a regime with a dynamically generated superpotential.
 By considering a simplified version of these theories we argue that
supersymmetry is probably broken.

\bigskip

\footnotetext[1]{Supported in part by DOE under cooperative 
                 agreement \#DE-FC02-94ER40818.}
\footnotetext[2]{NSF Young Investigator Award, Alfred P.~Sloan
Foundation Fellowship, DOE Outstanding Junior Investigator Award. }

\end{titlepage}

\section{Introduction}

After a lull of about ten years, the number of known models which
dynamically break supersymmetry has been steadily rising.  One begins
to suspect that the restricted number of theories was primarily due to a
limited ability to analyze strongly interacting theories. With recent
advances in understanding these theories~\cite{Seiberg,duality},
progress is being made in exploring the larger class of theories which can
break supersymmetry, leading to several new models of supersymmetry
breaking~\cite{dnns,Poppitz,it,yan}. A second problem with the search
for supersymmetry breaking is that theories with even a slightly
complicated field content can quickly become cumbersome to analyze.
This second problem can still be a frustration.

In this paper, we present an interesting nontrivial application of
exact methods to analyze a model which spontaneously breaks supersymmetry.
The theories that we analyze are based on the gauge group
$SU(n)\times SU(3) \times U(1)$. Because the gauge dynamics
are very different for $n=4$ and $n>4$, we first consider
the gauge group $SU(4) \times SU(3) \times U(1)$. The particular
models we explore in this paper are based on an idea discussed
in Ref.~\cite{dnns}, where it was suggested to search for models which
dynamically break supersymmetry by taking a known model and removing
generators to reduce the gauge group. This method is guaranteed to generate
an anomaly free chiral theory which has the potential to break supersymmetry.
There are several known examples of theories with a suitable superpotential
respecting the less restrictive gauge symmetries of the resultant theory,
in which supersymmetry is broken without runaway directions. However, there
is as yet no proof that this method will necessarily be successful. 

The $SU(n+3)$ theories for even $n$ with an antisymmetric tensor and $n-1$
antifundamentals are known to break supersymmetry
dynamically~\cite{SUasymmetric}. In this paper we consider models based on
the reduced gauge group $SU(n) \times SU(3) \times U(1)$.

Unlike previous models in the literature, neither of the nonabelian gauge
groups generates a dynamical superpotential in the absence of the
perturbations added at tree level. Because neither factor generates
a dynamical superpotential, there is no limit in which the theory can be
analyzed perturbatively. Therefore, we derive the exact superpotential
for the $n=4$ case which we use to show supersymmetry is broken in the
strongly interacting theory.

The $SU(4)\times SU(3) \times U(1)$  model is interesting for several
reasons. First, the demonstration of supersymmetry breaking involves
a subtle interplay between the confining dynamics and the tree-level
superpotential of the theory. Second, this model implements the mechanism
of~\cite{it,yan} without introducing additional singlets or potential runaway
directions. Third, we can lift all the flat directions by a renormalizable
superpotential. Fourth, none of the gauge groups generates a dynamical
superpotential; the fields are kept from the origin solely by a quantum
modified constraint.

In addition, the exact superpotential exhibits several novel features.
First, fields with quantum numbers corresponding to classically vanishing
gauge invariant operators emerge, and play the role of Lagrange multipliers
for known constraints. Second, we find that classical constraints can be
modified not only by a constant, but by field dependent terms which vanish
in the classical limit. Third, fields which are independent in the classical
theory satisfy linear constraints in the quantum theory. By explicitly
substituting the solution to the equation of motion for these fields,
we show that quantum analogs of the classical constraints are still satisfied.

The $SU(n) \times SU(3) \times U(1)$ theories for $n>4$ are less tractable
but nonetheless very interesting. We show that it is possible to introduce
Yukawa couplings which lift all classical flat directions. We then consider
the low-energy limit of this theory. The  $SU(3)$ gauge group without the
perturbative superpotential is not confining. However, the $SU(n)$ confined
theory in the presence of Yukawa couplings induces masses  for
sufficiently many flavors that there is a dynamical superpotential
associated with both the $SU(3)$ and  $SU(n)$ dynamics.  This low-energy
superpotential depends non-trivially on both the strong dynamical scales
of the low-energy theory and the Yukawa couplings of the microscopic theory.
We consider  this model with and without Yukawa couplings which lift the baryon
flat directions. In the first case, the theory is too complicated to solve.
The form of the low-energy superpotential permitted by the symmetries is
nonetheless quite interesting in that  it mixes   the perturbative and strong
dynamics. In the second case, we can explicitly derive that supersymmetry
is broken. In either case, there is a spontaneously broken global $U(1)$
symmetry, so we conclude this theory probably breaks supersymmetry and has
no dangerous runaway directions when all required Yukawa couplings are
nonvanishing.

The outline of this paper is as follows. We first describe the $SU(4) \times
SU(3) \times U(1)$ model classically. In particular, we show that the model
has no classical flat directions. In Section 3, we analyze the quantum
mechanical theory in the strongly interacting regime.  In Section 4, we show
that the model breaks supersymmetry. In Section 5, we discuss
generalizations to $SU(n) \times SU(3) \times U(1)$ and conclude in the final
section.

\section{The Classical $SU(4) \times SU(3) \times U(1)$ Theory}

The field content of the model we study is obtained by decomposing the
chiral multiplets of an $SU(7)$ theory with the field content consisting
of an antisymmetric tensor and three anti-fundamentals into its
$SU(4)\times SU(3)\times U(1)$ subgroup. The fields are:
\begin{displaymath}
 A^{\alpha\beta}(6,1)_6, \,  \bar{Q}_a(1,\bar{3})_{-8}, \,
 T^{\alpha a}(4,3)_{-1}, \, \bar{F}_{\alpha I}(\bar{4},1)_{-3}, \,
 \bar{Q}_{a i}(1,\bar{3})_4,
\end{displaymath}
where $i,I=1,2,3$ are flavor indices, while Greek letters denote $SU(4)$
indices and Latin ones correspond to $SU(3)$. In this notation $(n,m)_q$
denotes a field that transforms as an $n$ under $SU(4)$, $m$ under $SU(3)$
and has $U(1)$ charge $q$. 

 We take the classical superpotential  to be
\begin{eqnarray}
  \label{eq:superpot}
  W_{cl}  &  = &  A^{\alpha\beta} \bar{F}_{\alpha 1} \bar{F}_{\beta 2} +
          T^{\alpha a} \bar{Q}_{a 1} \bar{F}_{\alpha 1} +
          T^{\alpha a} \bar{Q}_{a 2} \bar{F}_{\alpha 2} +
          T^{\alpha a} \bar{Q}_{a 3} \bar{F}_{\alpha 3} + \nonumber \\
      & & \bar{Q}_a \bar{Q}_{b 2} \bar{Q}_{c 1} \epsilon^{abc}.
\end{eqnarray}
We will show shortly that this superpotential lifts all D-flat directions.

{}From the fundamental fields we can construct operators which are invariant
under the gauge symmetries of the theory. We first list those which are
invariant under $SU(4)\times SU(3)$ and subsequently construct operators
which are also $U(1)$ invariant. Later on it will be important to distinguish
operators invariant under the  confining gauge groups but which carry
$U(1)$ charge.
 \begin{eqnarray}
\label{eq:classical}
  M_{iI}= & T^{\alpha a} \bar{Q}_{a i} \bar{F}_{\alpha I}        & 0   
  \nonumber \\
  M_{4I}= & T^{\alpha a} \bar{Q}_{a} \bar{F}_{\alpha I}          & -12 
  \nonumber \\
  X_{IJ}= & A^{\alpha\beta} \bar{F}_{\alpha I} \bar{F}_{\beta J} & 0  
  \nonumber \\
  X_{I4}= & \frac{1}{6} A^{\beta \alpha} \bar{F}_{\beta I} 
            \epsilon_{\alpha \gamma \delta \zeta} T^{\gamma a}
            T^{\delta b} T^{\zeta c} \epsilon_{a b c}            & 0   
  \nonumber \\
 {\rm Pf}A=& \epsilon_{\alpha \beta \gamma \delta} 
            A^{\alpha \beta} A^{\gamma \delta}                   & 12  \\
  Y_{ij}= & \epsilon_{\alpha \beta \gamma \delta} 
            A^{\alpha \beta} T^{\gamma a} \bar{Q}_{ai}
            T^{\delta b} \bar{Q}_{b j}                           & 12  
   \nonumber \\
  Y_{i4}= & \epsilon_{\alpha \beta \gamma \delta} 
            A^{\alpha \beta} T^{\gamma a} \bar{Q}_{ai}
            T^{\delta b} \bar{Q}_{b}                             & 0   
   \nonumber \\
  \bar{B}=& \frac{1}{6} \bar{F}_{\alpha I} \bar{F}_{\beta J} 
            \bar{F}_{\gamma K} \epsilon^{I J K}
            T^{\alpha a} T^{\beta b} 
            T^{\gamma c} \epsilon_{a b c}                        & -12 
  \nonumber \\
  \bar{b}^i=& -\frac{1}{2} \bar{Q}_a \bar{Q}_{b j} \bar{Q}_{c k}
            \epsilon^{i j k} \epsilon^{a b c}                    & 0   
   \nonumber \\
  \bar{b}^4=& \frac{1}{6} \bar{Q}_{a i} \bar{Q}_{b j}
            \bar{Q}_{c k} \epsilon^{i j k} \epsilon^{a b c}      & 12  
   \nonumber 
\end{eqnarray}
The right hand side column indicates the charges of the operators under
the $U(1)$ gauge group. All other $SU(4)\times SU(3)$ invariants can
be obtained as products of these operators. The classical constraints obeyed
by these fields are: 
\begin{eqnarray}
\label{chargedconst}
  & & 4 \, X_{I4}X_{JK}\epsilon^{IJK}-\bar{B} \, {\rm Pf}A=0 \nonumber \\
   & & \epsilon^{ijk} \epsilon^{IJK} 
     \left( {\rm Pf}A \, M_{iI} M_{jJ} M_{kK} - 6 Y_{ij} M_{kI} X_{JK} \right) 
   = 0 \nonumber \\
 & & \epsilon^{ijk} \epsilon^{IJK} 
     \left({\rm Pf}A \, M_{4I} M_{jJ} M_{kK} - 2 Y_{jk} M_{4I} X_{JK}
     + 4 Y_{j4} M_{kI} X_{JK} \right) = 0 \nonumber \\ 
  & & Y_{i4}\bar{b}^i=0 \nonumber \\
  & & \bar{B}\bar{b}^4-\frac{1}{6} \epsilon^{ijk}\epsilon^{IJK}M_{iI}M_{jJ}
      M_{kK}=0 \nonumber \\
  & & \bar{B}\epsilon^{kij}Y_{ij}-2 \, \epsilon^{kij}\epsilon^{IJK}
      M_{iI}M_{jJ}X_{K4}=0 \nonumber \\
  & & M_{4I}\bar{b}^4+M_{iI}\bar{b}^i=0 \nonumber \\
  & & \epsilon^{ijk}Y_{jk}M_{4I}+2 \, \epsilon^{ijk}M_{jI}Y_{k4}+
      4 \, X_{I4}\bar{b}^i=0 \nonumber \\
  & & \epsilon^{IJK}\epsilon^{ijk}M_{iI}M_{jJ}M_{4K}Y_{k4}=0.
\end{eqnarray}

The completely gauge invariant fields  can be formed by taking products of
the above $U(1)$ charged fields. However, most of these combinations turn
out to be products of other completely gauge invariant operators. As an
operator basis we can use the neutral fields from Eq.~\ref{eq:classical}
and $E_I = M_{4I} {\rm Pf} A$. These operators are subject to the following
classical constraints:
\begin{eqnarray}
\label{neutconst}
  & & \epsilon^{I J K} E_{J} M_{i K} \bar{b}^i  =  0 \nonumber \\
  & & Y_{i 4} \bar{b}^i  =  0 \nonumber \\
  & & \epsilon^{I J K} \epsilon^{i j k} M_{i I} M_{j J} E_{K} Y_{k4} =  0
        \nonumber \\
  & & \epsilon^{I J K} \epsilon^{i j k} M_{i I} M_{j J} Y_{k 4}
        M_{l K} \bar{b}^l  =  0
\end{eqnarray}
These constraints follow from Eq.~\ref{chargedconst}. We have omitted the
linear constraints following from Eq.~\ref{chargedconst} which define
additional unnecessary fields. These operators obeying the above constraints
parameterize the D-flat directions of the theory.

In terms of the invariants defined above we can express the superpotential as
\begin{equation}
  W_{cl}  = X_{1 2} + M_{1 1} + M_{2 2} + M_{3 3} + \bar{b}^3.
\end{equation}
We now show that this superpotential suffices to lift all $D$-flat directions. 
It is easiest to show this (using the results of Ref.~\cite{LutyTaylor})
by demonstrating that the holomorphic invariants which parameterize the
flat directions are all determined by the equations of motion (as opposed
to parameterizing the flat directions in terms of the fundamental fields).
If all holomorphic invariants are determined,  we can conclude that all
potential flat directions are lifted.  

We consider the equations of motion corresponding to the classical
superpotential of Eq.~\ref{eq:superpot}. The equation
$\frac{\partial W}{\partial A}$ sets $X_{12}$ to zero if we multiply by $A$.
Forming all gauge invariant combinations from 
$\frac{\partial W}{\partial \overline{Q}_{a i}}$ we obtain the following.
Multiplying $\frac{\partial W}{\partial \overline{Q}_{a 3}}$ by
$\bar{Q}_{a j}$ gives
\begin{displaymath}
  M_{j 3}=0,
\end{displaymath}
similarly for $\frac{\partial W}{\partial \overline{Q}_{a 1,2}}$ we obtain
\begin{eqnarray*}
  M_{1 2}=0 &  M_{2 2} + \bar{b}^3=0 & M_{3 2} - \bar{b}^2=0 \\
  M_{2 1}=0 &  M_{1 1} + \bar{b}^3=0 & M_{3 1} - \bar{b}^1=0.
\end{eqnarray*}
Next, we multiply the same equations by $ \epsilon_{abc} T^{\beta b}
T^{\gamma c} A^{\delta \rho} \epsilon_{\beta \gamma \delta \rho}$ to obtain
\begin{displaymath}
  X_{34}=0 \; \; Y_{24} + 2 X_{14}=0 \; \; Y_{14} - 2 X_{24}  =0. 
\end{displaymath}
Also, by multiplying $\frac{\partial W}{\partial \overline{Q}_{a i}}$ by 
$\bar{Q}_a {\rm Pf} A$ we get
\begin{displaymath}
  E_I=0.
\end{displaymath}
Next, from $\frac{\partial W}{\partial \overline{Q}_{a }} \bar{Q}_{a }$
we obtain that
\begin{displaymath}
\bar{b}^3=0.
\end{displaymath}
We obtain the remaining equations from  $\frac{\partial W}{\partial 
\overline{F}_{\alpha I}}$. They are:
\begin{eqnarray*}
 M_{13} - X_{23}=0 &  M_{23} + X_{13}=0 & M_{3I}=0 \\
 E_2+ 4 Y_{14}=0   &  E_1- 4Y_{24}=0    & Y_{34}=0 
\end{eqnarray*}
The only solution to these equations sets all operators to be zero.
Therefore, our theory does not have flat directions.

 In Ref.~\cite{ADS2} it was argued that theories which have no flat directions,
but preserve an anomaly free $R$ symmetry break supersymmetry spontaneously
if the $U(1)_R$ symmetry is spontaneously broken in the vacuum. This follows
because there would be a massless pseudoscalar, which is unlikely to have 
a massless scalar partner. The superpotential of Eq.~\ref{eq:superpot}
preserves an $R$ symmetry under which the $R$ charges are
$R(A)=R(\bar{F}_3)=0$, $R(\bar{F}_1)=R(\bar{F}_2)=1$,
$R(\bar{Q}_1)=R(\bar{Q}_2)=\frac{5}{3}$, $R(\bar{Q}_3)=\frac{8}{3}$,
$R(\bar{Q})=-\frac{4}{3}$ and $R(T)=-\frac{2}{3}$. Although this symmetry
is anomalous with respect to the $U(1)$ gauge group, if it is spontaneously
broken, the associated Goldstone boson is nonetheless massless so the
argument of Ref.~\cite{ADS2}  should still   apply. 

Notice that the classical equations of motion in our theory have a solution
only where all fields vanish. In the next section we show that the quantum
 theory does not permit such a supersymmetric solution,
so that supersymmetry is broken.

\section{The Quantum $SU(4) \times SU(3) \times U(1)$ Theory}

In this section we will derive the exact superpotential of the
$SU(4)\times SU(3)\times U(1)$ theory. 
 The fact that it is possible to determine the exact superpotential
of the theory will enable us to  prove that supersymmetry is dynamically 
broken.

Before proceeding,  we list the global symmetries of the microscopic fields,
which are useful when constraining the form of the exact superpotential.
The global symmetries are:
\begin{displaymath}
\begin{array}{c|cccccccc} 
 & U(1)_A & U(1)_{\bar{Q}}& U(1)_T&U(1)_{\bar{F}}&
  SU(3)_{\bar{F}_I}&U(1)_{\bar{Q}_i}&SU(3)_{\bar{Q}_i}&U(1)_R \\ \hline
 A & 1&0&0&0&1&0&1&0\\
 \bar{Q}&0&1&0&0&1&0&1&0\\
 T&0&0&1&0&1&0&1&0\\
 \bar{F}_I&0&0&0&1&3&0&1&0\\
 \bar{Q}_i&0&0&0&0&1&1&3&0\\
 \Lambda_3^5&0&1&4&0&1&3&1&-2\\
 \Lambda_4^8&2&0&3&3&1&0&1&0\end{array}
\end{displaymath}
The only invariants under all global symmetries including $U(1)_R$ are
${\cal A}=X_{IJ} X_{K4} \epsilon^{IJK}/\Lambda_4^8$ 
and ${\cal B}=\bar{B}{\rm Pf}A/\Lambda_4^8$.

We now identify  the proper degrees of freedom. To do so, it is convenient
to first take the limit $\Lambda_3 \gg \Lambda_4$ and construct $SU(3)$
invariant operators which are mesons and baryons formed from the $SU(3)$ 
charged fields, and then to construct the $SU(4)$ bound states of these
fields. This gives us the spectrum which matches anomalies of the original
microscopic theory, independent  of the ratio $\Lambda_3/\Lambda_4$.

Below the $SU(3)$ scale, the theory can be described by an $SU(4)$ theory
with an antisymmetric tensor and four flavors. These four flavors are
\begin{eqnarray}
\label{su3invariants}
 & & \bar{F}_{\alpha 4}= \frac{1}{6} \epsilon_{\beta \gamma \delta \alpha}
     T^{\beta a}T^{\gamma b}T^{\delta c}\epsilon_{abc}, \nonumber \\
 & & F^{\alpha}_i=T^{\alpha a}\bar{Q}_{a i}, \; i=1,2,3 \nonumber \\
 & & F^{\alpha}_4=T^{\alpha a}\bar{Q}_{a},
\end{eqnarray}
The three remaining antifundamentals are $\bar{F}_{\alpha  I}$, $I=1,2,3$,
the original fields. The $SU(3)$ antibaryons are the $\bar{b}^i$'s
of Eq.~\ref{eq:classical}, which are singlets under $SU(4)$.

The four-flavor theory with an antisymmetric tensor has been described in
Ref.~\cite{Pouliot}. The confined states of the $SU(4)$ theory are
\begin{eqnarray}
& & {\rm Pf}A=\epsilon_{\alpha \beta \gamma \delta}A^{\alpha \beta}
    A^{\gamma \delta} \nonumber \\
& & M_{iI}=F^{\alpha}_i\bar{F}_{\alpha I} \nonumber \\
& & X_{IJ}=A^{\alpha \beta} \bar{F}_{\alpha I}\bar{F}_{\beta J} \nonumber \\
& & Y_{ij}=A^{\alpha \beta} F^{\gamma}_i F^{\delta}_j 
    \epsilon_{\alpha \beta \gamma \delta} \nonumber \\
& & B=\frac{1}{24} F^{\alpha}_i F^{\beta}_j F^{\gamma}_k F^{\delta}_l
    \epsilon_{\alpha \beta \gamma \delta}\epsilon^{ijkl} \nonumber \\
& & \bar{B}=\frac{1}{24} \bar{F}_{\alpha I} \bar{F}_{\beta J} 
\bar{F}_{\gamma K} \bar{F}_{\delta L}
\epsilon^{\alpha \beta \gamma \delta}\epsilon^{IJKL}. 
\end{eqnarray}
Here the indices $i$ and $I$ range from 1 to 4.
Note that $B, M_{44}$ and $M_{i4}$ are fields which vanish classically.
However, anomaly matching of the microscopic theory to the low-energy
theory requires the presence of these fields. Fields other than $B, M_{44}$
and $M_{i4}$  correspond to operators introduced in 
Eq.~\ref{eq:classical}.  The low-energy theory consists of the
fields listed in Eq.~\ref{eq:classical} and the new fields $B, M_{44}$,
and $M_{i4}$.

In order to construct the superpotential it is again convenient to consider 
the limit $\Lambda_3 \gg \Lambda_4$. Below the $\Lambda_3$ scale,  there is
an $SU(4)$ theory with four flavors and an antisymmetric tensor together
with the confining $SU(3)$ superpotential of Ref.~\cite{Seiberg}. The
superpotential for the four-flavor $SU(4)$ theory with an antisymmetric
tensor has been described in Ref.~\cite{Pouliot}. We determined the
coefficients in the superpotential of Ref.~\cite{Pouliot} by requiring that
the equations of motion reproduce the classical constraints.

In this limit, the superpotential has to be the sum of the contributions 
from $SU(3)$ and $SU(4)$ dynamics.  The exact superpotential  is therefore of
the form:
\begin{eqnarray}
\label{exsuppot}
W & \hspace{-4pt} = & \hspace{-4pt} \bar{b}^3 +X_{12} + M_{11} + M_{22}
                    + M_{33} + \frac{1}{\Lambda_3^5} \left( M_{i4} \bar{b}^i 
                    - B \right) + \nonumber \\
 \hspace{-4pt}& & \hspace{-4pt} f({\cal A},{\cal B}) \cdot
                  \frac{1}{24 \, \Lambda_3^5\Lambda_4^8} 
                  \Big( 24 \, BX_{IJ}X_{KL}\epsilon^{IJKL}+ 
                  6 \, \bar{B}Y_{ij}Y_{kl}\epsilon^{ijkl}
                  -24 \, B\bar{B}{\rm Pf}A + \nonumber \\
 \hspace{-4pt}& & \hspace{-4pt} 
     {\rm Pf}A \epsilon^{ijkl} \epsilon^{IJKL} M_{iI} M_{jJ} M_{kK} M_{lL}
     - 12 \, \epsilon^{ijkl} Y_{ij} M_{kI} M_{lJ} X_{KL}\epsilon^{IJKL} \Big),
\end{eqnarray}
where $f$ is an as yet undetermined function of the symmetry invariants
${\cal A}$ and ${\cal B}$, and $i,I=1,\ldots,4$.
Therefore, the symmetries together with the limit $\Lambda_3 \gg \Lambda_4$
restrict the  superpotential up to  a function of 
${\cal A}$ and ${\cal B}$. However, a negative power series in ${\cal A}$
or ${\cal B}$ would imply unphysical singularities, since there
is no limit in which the number of flavors in the $SU(4)$ theory
is less than the number of colors. On the other hand, a positive power
series in ${\cal A}$ or ${\cal B}$ would not correctly reproduce the limit
where $\Lambda_4 \gg \Lambda_3$. In this limit one has an $SU(4)$ theory
with an antisymmetric tensor and three flavors, which yields a quantum
modified constraint~\cite{Poppitz}. Observe the amazing fact that the $B$
equation of motion which involves the superpotential from both the $SU(3)$
and $SU(4)$ terms exactly reproduces this $SU(4)$ quantum modified constraint.
This is only true with no further modification of the second term. In fact,
this is what permits us to fix the relative coefficient of the two terms in
parentheses. Thus we conclude that $f({\cal A},{\cal B}) \equiv 1$.

We stress again that each of the fields $B$, $M_{i4}$, and $M_{44}$ vanish
classically. In the quantum theory, the $B$ field acts as a Lagrange
multiplier for the three flavor $SU(4)$ quantum modified constraint.
The $M_{i4}$ and $M_{44}$ equations of motion are  
\begin{eqnarray} \label{eomb}
 & & \epsilon^{ijk} \epsilon^{IJK} 
     \left( {\rm Pf}A M_{iI} M_{jJ} M_{kK} - 6 Y_{ij} M_{kI} X_{JK} \right) 
   = 6 \Lambda_4^8 \, \bar{b}^4 \\
 & & \epsilon^{ijk} \epsilon^{IJK} 
     \left({\rm Pf}A M_{4I} M_{jJ} M_{kK} - 2 Y_{jk} M_{4I} X_{JK}
     + 4 Y_{j4} M_{kI} X_{JK} \right) = 2 \Lambda_4^8 \, \bar{b}^i 
     \nonumber 
\end{eqnarray}
The linear equations for $\bar{b}^i$ and $\bar{b}^4$ can be understood
by the fact that they appear as mass terms for $M_{44}$ and $M_{i4}$.
The equations of motion in Eq.~\ref{eomb} can be interpreted as quantum
modified constraints of a three flavor $SU(4)$ theory with the scales
related through the $\bar{b}$-dependent masses.  

It is a nontrivial check on the superpotential of Eq.~\ref{exsuppot} that
all classical constraints have a quantum analog and vice versa.
The quantum modified constraints involving $\bar{b}^i$ and $\bar{b}^4$
are derived by substituting in the solution to their equation of motion.
The quantum modified constraints are:
\begin{eqnarray}
\label{qc1}
  & & \hspace{-1cm} 4 \, X_{I4}X_{JK}\epsilon^{IJK}-\bar{B}{\rm Pf}A=
      \Lambda_4^8  \\
   \label{qc2}
  & & \hspace{-1cm} \epsilon^{ijk} \epsilon^{IJK} 
     \left( {\rm Pf}A M_{iI} M_{jJ} M_{kK} - 6 Y_{ij} M_{kI} X_{JK} \right) 
   = 6 \Lambda_4^8 \, \bar{b}^4  \\
\label{qc3}
  & & \hspace{-1cm} \epsilon^{ijk} \epsilon^{IJK} 
     \left({\rm Pf}A M_{4I} M_{jJ} M_{kK} - 2 Y_{jk} M_{4I} X_{JK}
     + 4 Y_{j4} M_{kI} X_{JK} \right) = 2 \Lambda_4^8 \, \bar{b}^i  \\
\label{qc4}
  & & \hspace{-1cm} \epsilon^{IJK}\epsilon^{ijk}M_{iI}M_{jJ}M_{4K}Y_{k4}=
      2 \, B M_{4I} X_{JK} \epsilon^{IJK}  \\
\label{qc5}
  & & \hspace{-1cm} \bar{B}\epsilon^{kij}Y_{ij}-2 \, \epsilon^{kij}
      \epsilon^{IJK}
      M_{iI}M_{jJ}X_{K4}=-2\,M_{i4}M_{jI}\epsilon^{kij}X_{JK} \epsilon^{IJK}
\end{eqnarray}
while the remaining constraints are not modified. The interesting thing to
observe in the above equations is that the quantum modifications do not
simply involve addition  of a constant to the classical field equations.
The quantum modification can be field dependent. The classical limit is
recovered in Eqs.~\ref{qc4}, \ref{qc5} because $B$ and $M_{i4}$ are fields
which vanish classically. Without a tree-level superpotential $M_{i4}$
is set to zero by the $\bar{b}^i$ equations of motion. However, $M_{i4}$
can be non-vanishing in the presence of a tree-level superpotential. The
quantum modifications in Eqs.~\ref{qc2}, \ref{qc3} do not contain classically
vanishing fields, but are proportional to $\Lambda_4$, which ensures the
correct classical limit. This field dependent modification of constraints
is a new feature which is not present when analyzing simple nonabelian
gauge groups.

Note that five of our constraints (Eqs.~\ref{qc1}, \ref{qc2} and \ref{qc3})
can be interpreted as the quantum modified constraints on the moduli space
of an $SU(4)$ gauge theory with an antisymmetric tensor and three flavors.
Such a theory is obtained in several limits. If $\Lambda_4 \gg \Lambda_3$ 
one trivially has a three flavor $SU(4)$ theory with an antisymmetric tensor.
On the other hand, if $\Lambda_3 \gg \Lambda_4$ and any single $\bar{b}$
is non-vanishing one also has a three flavor $SU(4)$ theory with its
corresponding quantum modified constraint.

When deriving the constraints in Eqs.~\ref{qc1}-\ref{qc5} from the exact
superpotential we frequently encounter expressions containing inverse
powers of $\Lambda_4$. Such terms are singular in the limit when $\Lambda_3$
is held fixed and $\Lambda_4 \rightarrow 0$. This is true even for expressions
containing the fields $B, M_{i4}$ and $M_{44}$, since they vanish only
in the limit when $\Lambda_3 \rightarrow 0$. Therefore all such terms  
must and do cancel.

\section{Dynamical Supersymmetry Breaking}

In the low-energy description of our model the $SU(4)$ and $SU(3)$ gauge
groups are confined and the only remaining gauge group is the $U(1)$.
This $U(1)$ does not play any role in supersymmetry breaking; its purpose
is to lift some classical flat directions.  Unlike previous examples of
dynamical supersymmetry breaking, the superpotential can be completely
analyzed in a regime where there are no singularities, either due to a
dynamically generated superpotential present in the initial theory,
integrating out fields, or particular limits. If the theory breaks
supersymmetry, it is simply of O'Raifeartaigh type~\cite{OR}. In this
section, we show that this is the case; there is no consistent solution of 
the $F$-flatness equations for the exact superpotential of Eq.~\ref{exsuppot}.

 We first assume that $\bar{B}\neq 0$. Then  the 
$\frac{\partial W}{\partial Y_{ij}}$ equation of motion implies  
\begin{equation}
\label{yeq}
Y_{ij}=\frac{1}{\bar{B}}X_{KL}M_{iI} M_{jJ} \epsilon^{IJKL}.
\end{equation}
Plugging this expression into the $\frac{\partial W}{\partial X_{IJ}}$
equation of motion, we obtain 
\begin{displaymath}
  (\delta^3_S \delta^4_T-\delta^3_T \delta^4_S) + 
  \frac{8}{\Lambda_3^5 \Lambda_4^8} B X_{ST} - 
  \frac{2}{\Lambda_3^5 \Lambda_4^8} \frac{1}{\bar{B}} \epsilon^{ijkl}
  M_{iM} M_{jN} M_{kS} M_{lT} X_{KL} \epsilon^{MNKL}=0.  
\end{displaymath}
However, by using the $\frac{\partial W}{\partial {\rm Pf}A}=0$ equation
in the above expression we arrive at a contradiction. 

Next we assume that $\bar{B}=0$, but $B\neq 0$. We can
now solve for $X$ using the equation $\frac{\partial W}{\partial X_{IJ}}=0$:
\begin{equation}
\label{X}
  X_{MN}=\frac{\Lambda_3^5\Lambda_4^8}{8 B}
  \left[ (\delta_M^3\delta_N^4-\delta_N^3\delta_M^4)+
  48 \, \epsilon^{ijkl} Y_{ij} M_{kM} M_{lN} \right].
\end{equation}
Then we multiply this equation by $\epsilon^{ijkl}\epsilon^{IJMN}
M_{kI} M_{lJ}$. The $Y_{ij}$ equation of motion sets the left hand side to
zero, while the ${\rm Pf}A$ equation of motion sets the second term on
the right hand side to zero. Therefore,  
\begin{displaymath}
  \epsilon^{ijkl} M_{iI} M_{jJ} \epsilon^{IJ34}=0.
\end{displaymath}
Using this fact, the ${\rm Pf}A$ equation of motion, and the expression
for $X_{MN}$ in Eq.~\ref{X} we get that $\frac{\partial W}{\partial B}=
-\frac{1}{\Lambda_3^5}$, which again means that the equations
of motion are contradictory.

Finally we assume that $B=\bar{B}=0$. Then  the 
$\frac{\partial W}{\partial X_{IJ}}$ equation of motion  implies 
\begin{displaymath}
  \epsilon^{ijkl} Y_{ij} M_{kI} M_{lJ}=0
\end{displaymath}
for all $I,J$ except $I=3,J=4$. Multiplying the 
$\frac{\partial W}{\partial X_{IJ}}$ equation of motion by $M_{iI} M_{jJ}$
and using the $\frac{\partial W}{\partial {\rm Pf}A}$ equation of motion
we get that
\begin{displaymath}
  \epsilon^{ijkl} M_{i1} M_{j2}=0.
\end{displaymath}
Using these results the $\frac{\partial W}{\partial M_{i3}}$ equation of
motion yields
\begin{displaymath}
  \delta^{i3}-\frac{1}{\Lambda_3^5\Lambda_4^8} \epsilon^{ijkl}
  Y_{jk} M_{lJ} X_{KL}\epsilon^{3JKL}=0.
\end{displaymath}
Multiplying this equation by $M_{i4}$  implies $M_{34}=0$, which is in
contradiction with the $\frac{\partial W}{\partial \overline{b}^3}$ equation
of motion. Thus we have shown that this $SU(4)\times SU(3)\times U(1)$ model
breaks supersymmetry dynamically. Since there are no classical flat directions,
there should not be runaway directions in this model.

Having presented a general proof of supersymmetry breaking, we now give
a simpler proof that applies only in a restricted region of parameter space.
Assume that $\Lambda_3$ is the largest parameter in the theory. The effective
superpotential just below the $\Lambda_3$ scale is
\begin{eqnarray}
W&=& \bar{b}^3 + \gamma A^{\alpha \beta} \bar{F}_{\alpha 1} \bar{F}_{\beta 2}
   + \lambda_1 F^\alpha_1 \bar{F}_{\alpha 1} + 
     \lambda_2 F^\alpha_2 \bar{F}_{\alpha 2} +
     \lambda_3 F^\alpha_3 \bar{F}_{\alpha 3} + \nonumber \\
 & & \frac{1}{\Lambda_3^5} \left( \bar{F}_{\alpha 4} F^\alpha_i \bar{b}^i
     - {\rm det} F^\alpha_i \right),
\end{eqnarray}
where we use the notation from Eq.~\ref{su3invariants} and we introduced
explicitly the Yukawa couplings $\gamma$ and $\lambda_{1,2,3}$. In terms 
of the canonically normalized fields, $\lambda_{1,2,3}$ are mass parameters.

Next, we integrate out three of the four flavors to arrive at an $SU(4)$
theory with one flavor and a superpotential
\begin{equation}
  W=\bar{b}^3 +\frac{1}{\Lambda_3^5} \bar{F}_{\alpha 4} F^\alpha_4
    \bar{b}^4.  
\end{equation}
To describe the dynamics of the one-flavor $SU(4)$ theory, it is useful
to define the effective one-flavor $SU(4)$ scale $\tilde{\Lambda}_4^5$,
which is proportional to 
   $\lambda_1 \lambda_2 \lambda_3 \Lambda_3^5 \Lambda_4^8$.
Below the effective $\tilde{\Lambda}_4$ scale there is a dynamically
generated term, so the low-energy superpotential is 
\begin{equation}
  W=\bar{b}^3+\frac{1}{\Lambda_3^5} M_{44} \bar{b}^4 + 
    \left( \frac{\tilde{\Lambda}_4^5}{{\rm Pf}A \, M_{44}}\right)^\frac{1}{2},
\end{equation}
where $M_{44}=\bar{F}_{\alpha 4} F_4^\alpha$.
There are no solutions to the equations of motion. Note that the potential
runaway direction is removed by the $U(1)$ D-flatness condition. Therefore
supersymmetry is dynamically broken. Observe that supersymmetry breaking 
in this limit has two sources. First the 
superpotential generated by the $SU(3)$ and $SU(4)$ gauge groups together
does not have a supersymmetric minimum.
Second, a Yukawa term in the tree level superpotential 
is confined into a single field which is also a source of
supersymmetry breaking. In fact, the tree-level Yukawa terms have three
different important roles in this analysis. They lift the flat directions,
they yield mass terms for the $SU(4)$ fields after $SU(3)$ is confining,
and they also contribute to supersymmetry breaking by the linear term.
The fact that there is a quantum modified constraint in the
$\Lambda_4\gg\Lambda_3$ limit of the theory does not seem to play a major 
role in the dynamics of supersymmetry breaking.

By symmetries, it can be shown that this
simpler proof neglects power corrections proportional to
\begin{displaymath}
\left( \frac{\gamma^2 \bar{b}^i {\rm Pf}A \, M_{44}}{\lambda^4 
             (\Lambda_3^5)^2} \right)^k.
\end{displaymath}
This reflects the fact that here we are studying the effective theory
treating $\Lambda_3$ as large. The $\bar{b}^4$ equation of motion together
with the fact that there are no flat directions imply broken supersymmetry
even with these corrections incorporated.

\section{$SU(n) \times SU(3) \times U(1)$ Theories}

In this section we generalize the $SU(4)\times SU(3) \times U(1)$
model to  $SU(n)\times SU(3) \times U(1)$, with $n$ even. There are 
several interesting features of the dynamics of these theories. 
Without a tree-level superpotential the $SU(3)$ group is not confining.
However, the Yukawa couplings of the tree-level superpotential become mass
terms when the $SU(n)$ group confines. These mass terms drive the $SU(3)$
group into the confining regime as well. Confinement can change chiral
theories into non-chiral ones. In this example Yukawa couplings become mass
terms. In fact, the quantum modified constraint associated with the $SU(n)$
group of the initial theory does not appear to play an essential role in the
dynamics of supersymmetry breaking. Another interesting phenomena is that even
if we remove some of the couplings from the superpotential, so that some flat
directions are not lifted, these directions turn out to be lifted in the
quantum theory. In particular, once the Yukawa couplings turn into mass terms,
the $SU(3)$ antibaryon directions are automatically lifted.
  
As in Section 2, we obtain the field content for these models by decomposing
the fields of the $SU(n+3)$ theory with an antisymmetric tensor and $n-1$
anti-fundamentals to
$SU(n)\times SU(3) \times U(1)$:
\begin{eqnarray}
\Yasymm & \rightarrow & A^{\alpha\beta}(\Yasymm,1)_6 + 
                       \bar{Q}_a(1,\bar{3})_{-2 n} +
                       T^{\alpha a}(\Yfund,3)_{3-n} \nonumber \\
(n-1) \, \overline{\Yfund} & \rightarrow & \bar{F}_{\alpha I} 
       (\overline{\Yfund},1)_{-3} + \bar{Q}_{ai} (1,\bar{3})_n,
\end{eqnarray}
where $i,I=1,\ldots,n-1$.

In analogy to the 4-3-1 case, $SU(n)\times SU(3) \times U(1)$ invariants
are:
\begin{eqnarray}
  M_{iI} & = & T^{\alpha a} \bar{Q}_{ai} \bar{F}_{\alpha I} \nonumber \\
  X_{IJ} &=& A^{\alpha \beta} \bar{F}_{\alpha I} \bar{F}_{\beta J} \nonumber \\
  X_I &=& \frac{1}{6} A^{\alpha_{n} \alpha_{n-1}} \ldots A^{\alpha_4 \beta}
          \bar{F}_{\beta I} \epsilon_{\alpha_{n} \ldots \alpha_1}
          T^{\alpha_3 a} T^{\alpha_2 b}  T^{\alpha_1 c} \epsilon_{abc} 
          \nonumber \\
  Y_i &=&  A^{\alpha_{n} \alpha_{n-1}} \ldots A^{\alpha_4 \alpha_3}
           T^{\alpha_2 a} \bar{Q}_{a i} T^{\alpha_1 b} \bar{Q}_b
          \nonumber \\
  \bar{b}_{ij} &=& \bar{Q}_a \bar{Q}_{b i} \bar{Q}_{c j} \epsilon^{a b c}
             \nonumber \\
  E_I &=& \epsilon_{\alpha_n \ldots \alpha_1} A^{\alpha_{n} \alpha_{n-1}}
          \ldots A^{\alpha_2 \alpha_1} T^{\beta a} \bar{Q}_a \bar{F}_{\beta I}
\end{eqnarray}
We consider the following superpotential:
\begin{eqnarray}
\label{npotential}
 W&=& X_{12} + X_{34} + \ldots + X_{n-3,n-2} +
      \bar{b}_{23} + \bar{b}_{45} + \ldots + \bar{b}_{n-2,1} + \nonumber \\
  & & M_{11} + M_{22} + \ldots + M_{n-1,n-1}.
\end{eqnarray}
Observe  the relative shifts in the indices between the $X$ and $\bar{b}$
operators. One can check that not all flat directions are removed without
such a shift in the indices.
  
To demonstrate that all flat directions are lifted, one can use the same
method as described in Section 2. In this example, we require looking not
only at linear equations in the flat direction fields, but also higher order
equations, in order to demonstrate that no flat directions remain
in the presence of the  tree-level superpotential above.

We first use the $\bar{Q}_i$ and $\bar{F}_i$ equations of motion (contracted
with $\bar{Q}_k$ and $\bar{F}_j$). One will then find potential flat
directions which are labeled by $i=1,3,5,\ldots,2 [n/4]-1$
with equal values of $X_{2j-1,(2j-1+i)||(n-2)}=\bar{b}_{2j,(2j+i)||(n-2)}$,
where $j=1,2,3,\ldots,(n-2)/2$ labels  nonvanishing $X$ and $\bar{b}$
fields which are equal along the flat direction. Here, by $[x]$
we denote the greatest integer less than $x$, while we define
$m||n \equiv 1 + (m-1)\  {\rm Mod}\ n$. There is another set of potential
flat directions of the form 
$X_{2j,(2j+i)||(n-2)}=\bar{b}_{2j-1,(2j-1+i)||(n-2)}$, where again
$j=1,2,3,\ldots,(n-2)/2$ and $i=1,3,5,\ldots,2 [n/4]-1$. In the case when
$n=4 k$ and $i=k$, two potential flat directions described above are equal
to each other, so they represent just one flat direction. Altogether, there
are $(n-2)/2$ potential flat directions. One of these flat directions is
lifted trivially by the $A$ equation of motion. To see that the remaining
flat directions are lifted requires obtaining quadratic equations
in the flat direction of fields by suitably contracting the $T$ equations
of motion. These equations can be shown to have only the trivial solution
where all fields vanish. We have verified this explicitly in the cases
$n=6,8,10,$ and $12$, but we expect this method to generalize. 

One can also verify that the superpotential above preserves two $U(1)$
symmetries, one of which is an $R$ symmetry which is anomalous only with
respect to the $U(1)$ gauge group. From the quantum modified constraint
it can be shown that at least one of these $U(1)$ symmetries is spontaneously
broken. Since the theory has no flat directions and spontaneously breaks
a $U(1)$ symmetry, we expect that supersymmetry is broken. 

There is a possibility however that in the strongly interacting regime
there is a point at which supersymmetry is restored. We now consider the
quantum theory and argue that it is likely that supersymmetry is broken.
  
Without a tree-level superpotential the $SU(3)$ group is not confining
for $n>4$ since $N_f > \frac{3}{2} N_c$. We choose to use fields
transforming under $SU(3)$ instead of the $SU(3)$ invariant operators. 
The D-flatness conditions can then be imposed explicitly. Although in
principle one could use holomorphic invariants to parameterize the D-flat
directions, the naive application of this method would lead to incorrect
results at points of the moduli space where these invariants
vanish~\cite{LisaErich}. Although with careful choice of holomorphic
invariants  this problem can be circumvented, in practice it is simpler
to use the charged fields when the gauge group is not confining.

The $SU(n)$ group has three flavors and an antisymmetric tensor. Therefore
$SU(n)$ is confining and gives rise to a quantum modified constraint as
described in Ref.~\cite{Poppitz}. The $SU(n)$ invariants are:
\begin{eqnarray}
\label{SUnvar}
  X_{IJ} &=& A^{\alpha \beta} \bar{F}_{\alpha I} \bar{F}_{\beta J} \nonumber \\
  m^a_I  &=& T^{\alpha a} \bar{F}_{\alpha I} \nonumber \\
  {\rm Pf} A &=& \epsilon_{\alpha_n \ldots \alpha_1} 
  A^{\alpha_{n} \alpha_{n-1}}
                 \ldots A^{\alpha_2 \alpha_1} \nonumber \\
  y_a &=& A^{\alpha_{n} \alpha_{n-1}} \ldots A^{\alpha_4 \alpha_3}
          \epsilon_{\alpha_n \ldots \alpha_1} 
          T^{\alpha_2 b} T^{\alpha_1 c} \epsilon_{a b c}
\end{eqnarray}
together with the fields $\bar{Q}_a$ and $\bar{Q}_{ai}$. 

The superpotential below the $\Lambda_n$ scale is 
\begin{eqnarray}
\label{SUnsuper}
W &= & \alpha^{12} X_{12} + \ldots + \alpha^{n-3,n-2} X_{n-3,n-2} + 
      \beta^{23} \bar{Q}_a \bar{Q}_{b 2} \bar{Q}_{c 3} \epsilon^{a b c} 
      +\ldots + \nonumber \\ 
  & & \beta^{n-2,1} \bar{Q}_a \bar{Q}_{b, n-2} \bar{Q}_{c 1} \epsilon^{abc}+
      \lambda^{11} m^a_1 \bar{Q}_{a 1} + \ldots +  \lambda^{n-1,n-1} m^a_{n-1} 
      \bar{Q}_{a, n-1} + \nonumber \\
  & & \eta \, \Big(\frac{n-2}{3n} \, \epsilon_{abc} m^a_{I_1} m^b_{I_2}
      m^c_{I_3} X_{I_4 I_5} \ldots X_{I_{n-2} I_{n-1}} 
      \epsilon^{I_1 \ldots I_{n-1}} {\rm Pf}A - \nonumber \\
  & & y_a m^a_{I_1} X_{I_2 I_3} \ldots X_{I_{n-2} I_{n-1}}
      \epsilon^{I_1 \ldots I_{n-1}} + \Lambda_n^{2n} \Big),
\end{eqnarray}
where $\eta$ is a Lagrange multiplier and we have  explicitly included the
coupling constants  in the tree-level superpotential. In terms of $SU(n)$
invariants, some of the terms in the above superpotential are just mass
terms for $(n-1)$ flavors of $SU(3)$, which drive $SU(3)$ into the confining
phase. In the presence of these perturbations, nonperturbative $SU(3)$
dynamics will generate a superpotential. 
Similar results are found in Ref. \cite{pst}. We stress again that in the
underlying theory these interactions are Yukawa couplings and not mass
terms.

To analyze the low-energy theory, we introduce an additional flavor of
$SU(n)$ with mass $\mu$. We do this because the $SU(n)$ quantum modified
constraint or equivalently anomaly matching shows that $SU(3)$ must be
broken below the scale $\Lambda_n$ in the original theory. With an additional
flavor, the origin of moduli space is permitted and $SU(3)$ can remain
unbroken. This permits us to derive the confining superpotential with two
massless $SU(3)$ flavors. Although the correct theory is only recovered 
in the limit $\mu \rightarrow \infty$, we will analyze the theory in the
regime $\mu < \Lambda_n$ and hope one can extrapolate the conclusion that
supersymmetry is broken~\cite{vecfl}. 

The superpotential with the additional massive $SU(n)$ flavor is:
\begin{eqnarray}
W&=& \alpha^{12} X_{12} + \ldots + \alpha^{n-3,n-2} X_{n-3,n-2} + \nonumber \\
 & &  \beta^{23} \bar{Q}_a \bar{Q}_{b 2} \bar{Q}_{c 3} \epsilon^{a b c} +
      \ldots + \beta^{n-2,1} \bar{Q}_a \bar{Q}_{b, n-2} \bar{Q}_{c 1} 
      \epsilon^{a b c} + \nonumber \\
 & &  \lambda^{11} m^a_1 \bar{Q}_{a 1} + \ldots +  \lambda^{n-1,n-1} m^a_{n-1}
      \bar{Q}_{a, n-1} + \mu m^4_n + \nonumber \\
 & &  \frac{1}{\Lambda_n^{2 n-1}} \Big( {\rm Pf}A \, m^a_{I_1} m^b_{I_2}
      m^c_{I_3} m^d_{I_4} X_{I_5 I_6} \ldots X_{I_{n-1} I_n}
      \epsilon_{abcd} \epsilon^{I_1 \ldots I_n} + \nonumber \\
 & &  Y^{a b} m^c_{I_1} m^d_{I_2} X_{I_3 I_4} \ldots X_{I_{n-1} I_n}
      \epsilon_{abcd} \epsilon^{I_1 \ldots I_n} +
      B X_{I_1 I_2} \ldots X_{I_{n-1} I_n} \epsilon^{I_1 \ldots I_n} + 
      \nonumber \\
 & &  \bar{B} Y^{ab} Y^{cd} \epsilon_{abcd} + B \bar{B} {\rm Pf}A \Big),    
\end{eqnarray}
where the variables are as defined in Eq.~\ref{SUnvar} with an extra $SU(n)$
flavor and 
\begin{eqnarray}
 B & = & T^{\alpha_1 a} T^{\alpha_2 b} T^{\alpha_3 c} F^{\alpha_4 4}
         A^{\alpha_5 \alpha_6} \ldots A^{\alpha_{n-1} \alpha_n} 
         \epsilon_{abc} \, \epsilon_{\alpha_1 \ldots \alpha_n} \nonumber \\
\bar{B}&=& \bar{F}_{\alpha_1 I_1} \ldots \bar{F}_{\alpha_n I_n}
           \epsilon^{I_1 \ldots I_n} \epsilon^{\alpha_1 \ldots \alpha_n}
           \nonumber \\
 Y^{a4}&=& T^{\alpha_1 a} F^{\alpha_2 4} A^{\alpha_3 \alpha_4} \ldots
          A^{\alpha_{n-1} \alpha_n} \epsilon_{\alpha_1 \ldots \alpha_n}
          \nonumber \\
 Y^{ab}&=& \epsilon^{abc} y_c.
\end{eqnarray}
The extra $SU(n)$ flavor is denoted by $F^{\alpha 4}$ and $\bar{F}_{\alpha n}$,
and $\Lambda_n$ is the dynamical scale of the four-flavor $SU(n)$ theory.
Here we have not bothered to establish the correct coefficients in the
last term in parentheses, since they are irrelevant in the forthcoming
analysis.

To arrive at the true low-energy theory, one would integrate out $n-3$
flavors, at which point a superpotential is generated involving $\Lambda_3$
for the four flavor theory. Upon integrating out the two remaining heavy
flavors, one would generate a complicated superpotential, involving both
the Yukawa couplings and the dynamical scales $\Lambda_n$ and $\Lambda_3$.
It is however technically difficult to explicitly perform this procedure
because of the nonlinear terms induced by the baryon operators in the
tree-level superpotential.  

If we instead constrain the form of the low-energy superpotential with
symmetries and limits, we find that the analysis remains quite complicated,
because many terms are permitted by the symmetries and physical limits. We
deduce the allowed terms by introducing a parameter $\tilde{\Lambda}_3$ which
transforms under anomalous global symmetries associated with the rotation of
each field carrying $SU(3)$ gauge charge in the initial microscopic theory.
Alternatively, we can define $\Lambda_3$ for the two flavor theory, where all
heavy flavors have been integrated out. The parameters 
$\tilde{\Lambda}_3^{9-n} {\rm det}(\lambda^{iI})/\Lambda_n^{2n-1}$
and $\Lambda_3^7$ have the same charge under all anomalous symmetries
so we can describe the low energy dynamics in terms of either one. 
We also see that if we consider $\tilde{\Lambda}_3$ as a fundamental finite
parameter of the initial theory, singularities in the Yukawa couplings
$\lambda^{iI}$ are permitted when we express the result in terms of the
low-energy $\Lambda_3 $, since the appropriate ratio is finite. In essence,
the Yukawa couplings become mass terms in the  $SU(n)$ confined theory, and
appear in the matching of $\Lambda_3$ across mass thresholds.

Examples of terms permitted by all symmetries and limits are:  
\begin{eqnarray}
 & & \frac{\Lambda_3^7}{\Lambda_n^{2n-1}} \, 
     \frac{\beta^{ij}}{(\lambda^{iI})^2}
     (X_{IJ})^{(n-4)/2} {\rm Pf}A  \, M^4_I \frac{1}{y_a Y^{a4}}, \nonumber \\
 & & \frac{\Lambda_3^{14}}{\Lambda_n^{2n-1}} \, 
         \frac{(\beta^{ij})^2}{(\lambda^{iI})^4}
  X_{In} (X_{IJ})^{(n-6)/2} {\rm Pf}A \frac{1}{(y_a Y^{a4})(y_a M^a_n)},
      \nonumber
\end{eqnarray}
where $\beta^{ij}$'s are the coefficients of the baryon operators $\bar{Q}
\bar{Q}_i \bar{Q}_j$, and $\lambda^{iI}$ of the $T \bar{F}_I \bar{Q}_i$ 
terms in the tree-level superpotential, but the index structure is not
specified. These terms mix the effects of the strong dynamics with the
tree-level superpotential, which is purely a consequence of integrating
out heavy fields. This does not violate the conjecture
of Refs.~\cite{Seiberg,KL}, which states that the couplings of the light
fields are not mixed into the dynamically generated superpotential.

Because of the complicated superpotential, the analysis of the full theory
is difficult. We will therefore consider a simpler version of the theory, in
which the baryon couplings, $\beta^{ij}$, are zero. This simplified
superpotential does not lift all flat directions classically, which might
lead to runaway directions in the quantum theory. One can show that these
remaining classical flat directions can be parameterized by the baryon
operators $\bar{b}_{ij}$. However, in the $SU(n)$ confined theory, these
fields are not flat, since the terms proportional to $m_{iI}$, which are
Yukawa couplings in the classical theory, are mass terms in the confined
theory. In this case, there is a potential for the baryon fields which drives
them towards the origin, and the baryon flat directions are lifted in the
quantum theory. This is similar in spirit to what was found in Ref.~\cite{it}.
In that example however, a quadratic constraint becomes a linear constraint
so the flat direction is removed; here we simply see that the $SU(n)$ confined
superpotential is such that the baryon fields are not flat. However there
is a caveat to this analysis which we discuss shortly.

In this limit it is simple to integrate out the heavy flavors and arrive
at the low-energy theory. The resulting superpotential is 
\begin{eqnarray}
W&=& \frac{1}{\Lambda_n^{2n-1}} \Big( y_a m^a_n m^4_{I_1} X_{I_2 I_3}
    \ldots X_{I_{n-2} I_{n-1}} \epsilon^{I_1 \ldots I_{n-1}} +
    B X_{I_1 I_2} \ldots X_{I_{n-1} I_n} \epsilon^{I_1 \ldots I_n} 
    \nonumber \\
& & + \bar{B} Y^{a4} y_a  + B \bar{B} {\rm Pf}A \Big) + \mu m^4_n +
     \lambda^{1 2} X_{1 2} + \ldots + \lambda^{n-3,n-2} X_{n-3,n-2}
     \nonumber \\
& & +\frac{\Lambda_3^7}{(Y^{a4} y_a)(m^b_n \bar{Q}_b) - 
    (Y^{a4} \bar{Q}_a)(m^b_n y_b)}.  
\end{eqnarray}
  This superpotential
clearly breaks supersymmetry since $m_n^4$ appears only in the term
$\mu m_n^4$. Since the scales of the $SU(n)$ theory with and without
extra flavor are related by $\mu \Lambda_n^{2n-1}=\Lambda_n^{2n}$,
this presumably implies that supersymmetry breaking is characterized
by $\Lambda_n^{2n-1}$ in the original theory.

Thus we just showed that if the $SU(n)$ gauge group is confining,
supersymmetry is broken. Had supersymmetry not been broken, this would
have been a good assumption, since all operators involving fields
transforming under the $SU(n)$ are driven to the origin by the classical
potential. Because supersymmetry is broken, it is conceivable that the true
vacuum is in the Higgs, rather than the confining phase. Nonetheless, we
still expect supersymmetry to be broken since there are no classically flat
directions in the theory. In this case however, the $\bar{b}$ operators are
not lifted by the superpotential. Once the effect of supersymmetry breaking
and the K\"ahler potential are included, the $\bar{b}$ fields presumably
have a nontrivial potential. We have not analyzed whether or not this can
give rise to runaway directions, should the Higgs phase prove to be the
true vacuum.
 
Having  argued that supersymmetry is probably broken for $\beta^{ij}=0$,
we hope that by including the remaining couplings, while lifting the flat
directions,  does not introduce a supersymmetric minimum. We expect that
the arguments presented above indicate that supersymmetry is broken in
the full $SU(n) \times SU(3) \times U(1)$ theories.

\section{Conclusions}
We have explored a new class of theories based on a product group, in which
neither gauge group  generates a dynamical superpotential in the absence
of perturbations. Nonetheless by exploring the exact superpotential, we
could explicitly demonstrate that supersymmetry is broken in the
$SU(4)\times SU(3)\times U(1)$ model. We also found interesting phenomena
in the exact superpotential, which were discussed in Section 3.  For the
$SU(n)\times SU(3)\times U(1)$ models, we have found that the exact
superpotential is quite complicated. However, in theories with
$\beta^{ij}=0$, we could demonstrate supersymmetry breaking with
the addition of an extra flavor of $SU(n)$. In this theory, we also
found a large number of classically flat directions
which are lifted in the quantum mechanical theory. This is due to the
fact that  when  $SU(n)$ confines, some of the Yukawa couplings in the 
tree-level superpotential turn into mass terms. This drives the $SU(3)$ group
into the confining region and also lifts some of the classical flat directions.
Although the particular example we studied in this paper involved
a gauge group which had a quantum modified constraint, this fact does not
seem essential to supersymmetry breaking in the $SU(n)\times SU(3)\times U(1)$
models, and the same mechanism should apply more generally.
 
That such interesting features appear in a fairly straightforward example
seems indicative of future possibilities. Although the classical theory is
constructed according to ``standard" rules, in that one can lift all flat
directions and spontaneously break an $R$ symmetry, the breaking of
supersymmetry is more subtle than in previous models. Verifying that
supersymmetry is broken in the full strongly interacting theory is
complicated because of the presence of many fields, even when the strong
dynamics is well understood.  It might be thought that the above properties
are sufficient for supersymmetry breaking; however it is not clear to us
that there cannot exist a point in the strongly interacting theory at which
supersymmetry is preserved.  Ultimately it would be interesting if it can
more rigorously be shown that models with the above properties necessarily
break supersymmetry.

Another intriguing observation is that the theories based on an existing
supersymmetric theory with generators removed from the original gauge group
with a sufficiently general superpotential seem to permit supersymmetry
breaking with no dangerous flat directions. In this paper, we have explored
an example distinct from previous ones in which the subgroup of the initial
gauge group is a  product group for which neither group generates a dynamical
superpotential.  We have shown that supersymmetry is broken in this case as
well, and presumably many other examples can be constructed along these lines
and analyzed with the full power of recent developments in strongly
interacting gauge theories.  It would be worthwhile to analyze these theories,
and also to see whether  it can be proven in general that theories constructed
in this fashion with a sufficiently general superpotential will break
supersymmetry without runaway directions.

We have not addressed the issue of the applicability of our models to
visible sector scenarios. In the $SU(4)\times SU(3) \times U(1)$ model, the
original theory  can preserve a global $SU(2)$ symmetry, and the
$SU(n)\times SU(3) \times U(1)$ model preserves a global $U(1)$ (in addition
to the $R$ symmetry). Since we have not analyzed the vacuum of our theories
in detail, we have not checked whether any of the global symmetries of the
classical theory were preserved by the supersymmetry breaking vacuum.
The $SU(n)\times SU(3)\times U(1)$ theories with $\beta^{ij}=0$ perhaps
suggest interesting possibilities, since there are many fields which seem
to play no role in supersymmetry breaking. There is a possibility that gauge
and/or global symmetries in this or similar models are left unbroken.
It might be possible to allow for more direct couplings between the
supersymmetry and visible sectors in this case.

\section*{Acknowledgments}
We are extremely  grateful to  Erich Poppitz, Rob Leigh and Martin Schmaltz
for their many helpful insights and suggestions. We also thank Daniel
Freedman, Philippe Pouliot, Riccardo Rattazzi, Nathan Seiberg and Yuri
Shirman for useful discussions. We thank Erich Poppitz, Yael Shadmi,
and Sandip Trivedi for sharing their results with us prior to publication.
LR thanks Rutgers University for its hospitality during the initial stages
of this project.

\end{document}